\documentclass[twocolumn]{aastex631}
\usepackage{graphics}
\usepackage{tabu}
\usepackage{natbib}
\usepackage{color}
\usepackage{rotating}
\usepackage{multirow}
\usepackage{footnote}
\usepackage{datetime}
\usepackage{amsmath}
\usepackage{mathrsfs}
\usepackage[normalem]{ulem}

\newcommand{\kms}{\ifmmode {\rm km\ s}^{-1} \else km s$^{-1}$\fi}
\newcommand{\lledd}{$L/L_{\rm Edd}$}
\newcommand{\lleddc}{$L/L_{\rm Edd,~corr}$}
\newcommand{\msun}{$M_{\odot}$}
\newcommand{\mbh}{$M_{\rm BH}$}
\newcommand{\mbhc}{$M_{\rm BH,~corr}$}
\newcommand{\cdist}{C\,{\sc iv}~$\parallel$~Distance}

\newcommand{\xray}{\hbox{X-ray}}

\newcommand{\hb}{H$\beta$}
\newcommand{\vLv}{$\nu L_{\nu} (5100$~\AA)}
\newcommand{\geminiN}{{\sl Gemini-North}}
\newcommand{\civ}{C\,{\sc iv}}
\newcommand{\fetwo}{Fe\,{\sc ii}}
\newcommand{\othree}{O\,{\sc iii}}
\newcommand{\magtwo}{Mg\,{\sc ii}}
\newcommand{\lya}{Ly$\alpha$}
\newcommand{\nv}{N\,{\sc v}}
\newcommand{\rfe}{$R_{\rm Fe\,II}$}

\graphicspath{{./}{figures/}}

\begin{document}

\title{Shedding New Light on Weak Emission-Line Quasars in the \civ --\hb~Parameter Space}
\author[0000-0001-6600-2517]{Trung Ha}
\affil{Department of Physics, University of North Texas, Denton, TX 76203, USA}
\email{trungha@my.unt.edu}

\author[0000-0003-0192-1840]{Cooper Dix}
\affil{Department of Physics, University of North Texas, Denton, TX 76203, USA}

\author[0000-0001-8406-4084]{Brandon M. Matthews}
\affil{Department of Physics, University of North Texas, Denton, TX 76203, USA}

\author[0000-0003-4327-1460]{Ohad Shemmer}
\affil{Department of Physics, University of North Texas, Denton, TX 76203, USA}

\author[0000-0002-1207-0909]{Michael S. Brotherton}
\affil{Department of Physics and Astronomy, University of Wyoming, Laramie, WY 82071, USA}

\author{Adam D. Myers}
\affil{Department of Physics and Astronomy, University of Wyoming, Laramie, WY 82071, USA}

\author[0000-0002-1061-1804]{Gordon T. Richards}
\affil{Department of Physics, Drexel University, 32 S 32nd St., Philadelphia, PA 19104, USA}

\author[0000-0002-4423-4584]{Jaya Maithil}
\affil{Center for Astrophysics $\vert$ Harvard \& Smithsonian, 60 Garden Street, Cambridge, MA 02138, USA}

\author[0000-0002-6404-9562]{Scott F. Anderson}
\affil{Department of Astronomy, University of Washington, Box 351580, Seattle, WA 98195, USA}

\author[0000-0002-0167-2453]{W. N. Brandt}
\affil{Department of Astronomy and Astrophysics, The Pennsylvania State University, University Park, PA 16802, USA}
\affil{Institute for Gravitation and the Cosmos, The Pennsylvania State University, University Park, PA 16802, USA}
\affil{Department of Physics, 104 Davey Lab, The Pennsylvania State University, University Park, PA 16802, USA}

\author{Aleksandar M. Diamond-Stanic}
\affil{Department of Physics and Astronomy, Bates College, Lewiston, ME, 04240, USA}

\author[0000-0003-3310-0131]{Xiaohui Fan}
\affil{Steward Observatory, University of Arizona, 933 North Cherry Avenue, Tucson, AZ 85721, USA}

\author[0000-0001-6217-8101]{S. C. Gallagher}
\affil{Department of Physics and Astronomy, University of Western Ontario, 1151 Richmond St, London, ON N6C 1T7, Canada}

\author[0000-0003-1245-5232]{Richard Green}
\affil{Steward Observatory, University of Arizona, 933 North Cherry Avenue, Tucson, AZ 85721, USA}

\author{Paulina Lira}
\affil{Departamento de Astronomía, Universidad de Chile, Casilla 36D, Santiago, Chile}

\author[0000-0002-9036-0063]{Bin Luo}
\affil{School of Astronomy and Space Science, Nanjing University, Nanjing, Jiangsu 210093, People's Republic of China}
\affil{Key Laboratory of Modern Astronomy and Astrophysics (Nanjing University), Ministry of Education, Nanjing, Jiangsu 210093, People's Republic of China}

\author[0000-0002-6766-0260]{Hagai Netzer}
\affil{School of Physics and Astronomy, Tel Aviv University, Tel Aviv 69978, Israel}

\author[0000-0002-7092-0326]{Richard M. Plotkin}
\affil{Department of Physics,University of Nevada, Reno, NV 89557, USA}
\affil{Nevada Center for Astrophysics, University of Nevada, Las Vegas, NV 89154, USA}

\author[0000-0001-8557-2822]{Jessie C. Runnoe}
\affil{Department of Physics and Astronomy, Vanderbilt University, Nashville, TN 37235, USA}

\author[0000-0001-7240-7449]{Donald P. Schneider}
\affil{Department of Astronomy and Astrophysics, The Pennsylvania State University, University Park, PA 16802, USA}
\affil{Institute for Gravitation and the Cosmos, The Pennsylvania State University, University Park, PA 16802, USA}

\author[0000-0002-0106-7755]{Michael A. Strauss}
\affil{Department of Astrophysical Sciences, Princeton University, Princeton, NJ 08544, USA}

\author[0000-0002-3683-7297]{Benny Trakhtenbrot}
\affil{School of Physics and Astronomy, Tel Aviv University, Tel Aviv 69978, Israel}

\author[0000-0001-7349-4695]{Jianfeng Wu}
\affil{Department of Astronomy, Xiamen University, Xiamen, Fujian 361005, People's Republic of China}

\received{December 1, 2022}
\revised{March 19, 2023}
\accepted{April 9, 2023}

\begin{abstract}

Weak emission-line quasars (WLQs) are a subset of Type~1 quasars that exhibit extremely weak \lya$+$\nv~$\lambda$1240 and/or \civ~$\lambda$1549 emission lines. We investigate the relationship between emission-line properties and accretion rate for a sample of 230 `ordinary' Type 1 quasars and 18 WLQs at $z < 0.5$~and $1.5 < z < 3.5$~that have rest-frame ultraviolet and optical spectral measurements. We apply a correction to the \hb-based black-hole mass (\mbh) estimates of these quasars using the strength of the optical \fetwo~emission. We confirm previous findings that WLQs' \mbh~values are overestimated by up to an order of magnitude using the traditional broad emission-line region size-luminosity relation. With this \mbh~correction, we find a significant correlation between \hb-based Eddington luminosity ratios and a combination of the rest-frame \civ~equivalent width and \civ~blueshift with respect to the systemic redshift. This correlation holds for both ordinary quasars and WLQs, which suggests that the two-dimensional \civ~parameter space can serve as an indicator of accretion rate in all Type~1 quasars across a wide range of spectral properties.

\end{abstract}

\keywords{galaxies: active --- quasars: emission line --- quasars: weak emission-line}

\section{Introduction}
\label{sec:intro} 
\setcounter{footnote}{0} 

Weak emission-line quasars (WLQs) are a subset of Active Galactic Nuclei (AGN) with extremely weak or undetectable rest-frame UV emission lines \citep[e.g.,][]{1999ApJ...526L..57F,2001AJ....122..503A,2005AJ....129.2542C,2010AJ....139..390P}. The Sloan Digital Sky Survey \citep[SDSS;][]{2000AJ....120.1579Y} has discovered $\approx 10^{3}$ Type~1 quasars with \lya$+$\nv~$\lambda1240$ rest-frame equivalent width (EW) $< 15.4$~\AA~and/or \civ~$\lambda1549$ EW $< 10.0$~\AA ~\citep[e.g.,][]{2009ApJ...699..782D,2014A&A...568A.114M}. These numbers represent a highly significant concentration of quasars at $\gtrsim 3 \sigma$ deviation from the log-normal EW distribution of the SDSS quasar population, with no corresponding ``tail" at the opposite end of the distribution \citep[][]{2009ApJ...699..782D, 2012ApJ...747...10W}. Furthermore, the fraction of WLQs among the broader quasar population increases sharply at higher redshifts (and thus higher luminosities), from $\sim 0.1\%$~at $3 \lesssim z \lesssim 5$~to $\sim 10-15\%$~at $z \gtrsim 5.7$ \citep[][]{2009ApJ...699..782D,2016ApJS..227...11B,2019ApJ...873...35S}. 

Multi-wavelength observations of sources of this class have shown that they are unlikely to be high-redshift galaxies with apparent quasar-like luminosity due to gravitational-lensing amplification, dust-obscured quasars, or broad-absorption-line (BAL) quasars \cite[e.g.,][]{2006ApJ...644...86S,Shemmer10}, but that their UV emission-lines are intrinsically weak. Furthermore, WLQs are typically radio-quiet, and have \xray~and mid-infrared properties inconsistent with those of BL Lac objects \citep{2009ApJ...696..580S,2011ApJ...743..163L,2012ApJ...747...10W,2017ApJ...834..113M}. 

About half of WLQs have notably lower \xray~luminosities than expected from their monochromatic luminosities at $2500$~\AA~\citep[e.g.,][]{2015ApJ...805..122L,2018MNRAS.480.5184N,2022MNRAS.511.5251N,2020MNRAS.492..719T}. One explanation for this phenomenon is that, at small radii, the geometrically thick accretion disks of these WLQs are `puffed up' and prevent highly ionizing photons from reaching the broad emission-line region \citep[BELR; e.g,][]{2011ApJ...736...28W,2012ApJ...747...10W,2015ApJ...805..122L,2018MNRAS.480.5184N,2022MNRAS.511.5251N}. The \xray~radiation is partially absorbed by the thick disk, resulting in low apparent \xray~luminosities at high inclinations (i.e., when these objects are viewed edge-on). When these objects are viewed at much lower inclinations, their notably steep \xray~spectra indicate accretion at high Eddington luminosity ratio \citep[$L_{\rm bol}/L_{\rm Edd}$, hereafter \lledd; e.g.,][]{2008ApJ...682...81S,2015ApJ...805..122L,Marlar18}.

The indications of high Eddington ratios in WLQs may provide a natural explanation for the weakness of their emission lines in the context of the Baldwin Effect. In its classical form, this effect is an anti-correlation between the EW(\civ) and the quasar luminosity \citep{1977ApJ...214..679B}. Subsequent studies, however, have found that this relation stems from a more fundamental anti-correlation between EW(\civ) and \hb-based \lledd~(\citeauthor{2004MNRAS.350L..31B} 2004, hereafter BL04;~\citeauthor{2009ApJ...703L...1D} 2009). This anti-correlation, coined the Modified Baldwin Effect (MBE), was extensively studied and built upon by \citet[][hereafter SL15]{Shemmer15}~\cite[however, see also][]{2022MNRAS.515.5836W}. SL15 utilized a sample of nine WLQs and 99 non-radio-loud, non-BAL (`ordinary') quasars spanning wide ranges of luminosity and redshift to analyze the relative strength of the \civ~emission-line and the \hb-based Eddington ratio. They confirmed the findings of BL04 for the sample of ordinary quasars. However, all nine WLQs were found to possess relatively low \lledd~values, while the MBE predicts considerably higher Eddington ratios for these sources. This finding led SL15 to conclude that the \hb-based \lledd~parameter cannot depend solely on EW(\civ)~for all quasars. Such a conclusion may also be consistent with subsequent findings that WLQs possess strong \fetwo~emission and large velocity offsets of the \civ~emission-line peak with respect to the systemic redshift (hereafter, Blueshift(\civ)) \citep{2018A&A...618A.179M}, and that \lledd~correlates with Blueshift(\civ) at high Blueshift(\civ) values \citep[see Figure 14 of][]{2020MNRAS.492.4553R}.

In this work, we explore two possible explanations for the findings of SL15. The first of these is that the traditional estimation of \hb-based black-hole mass (\mbh) values, and therefore \lledd~values, fails to accurately predict \mbh, particularly in quasars with strong optical \fetwo~emission \citep[e.g.,][]{2013BASI...41...61S,2022MNRAS.tmp.1709M}. Such a case is typical for WLQs, and thus a correction via measurement of the strength of the \fetwo~emission-complex in the optical band is required \citep[][]{2019ApJ...886...42D,2020MNRAS.491.5881Y}. The second explanation is that EW(\civ), by itself, is not an ideal indicator of the quasar accretion rate. In addition to EW(\civ), we utilize a recently defined parameter, the `\cdist' \citep[][hereafter R22]{2022ApJ...931..154R}, which represents a combination of the EW(\civ) and Blueshift(\civ) \citep{2011AJ....141..167R,2020ApJ...899...96R,CIVdist}, and search for a correlation between that parameter and \lledd. 

To investigate these explanations, we extend the WLQ sample of SL15 to nine additional sources available from the Gemini~Near-IR Spectrograph - Distant Quasar Survey \citep[GNIRS-DQS;][hereafter Paper~I]{M23}. Furthermore, we study the distribution of WLQs in  \lledd~space versus a sample of ordinary quasars from SL15 and Paper~I. We aim to investigate the underlying driver for the weak emission lines in WLQs and test the assertion that all WLQs have extremely high accretion rates.

The structure of this paper is as follows. In Section~\ref{sec:data}, we discuss our sample selection and the relevant equations used to estimate \hb-based \lledd~values. In Section~\ref{sec:results}, we analyze the samples' spectroscopic properties as well as the sources' black-hole masses and accretion rates. Subsequently, we discuss the correlation between the \civ~parameter space and \lledd. In Section~\ref{sec:conclusions}, we summarize our findings. Throughout this paper, we compute luminosity distances using a standard $\Lambda$CDM cosmology with $H_{\rm 0} = 70$~\kms~Mpc$^{-1}$, $\Omega_{\rm M} = 0.3$, and $\Omega_{\rm \Lambda} = 0.7$~\citep[e.g.,][]{Spergel07}.

\section{Sample Selection and Data Analysis}\label{sec:data}

\subsection{WLQ Sample}\label{sec:WLQs}

We compile a sample of 18 WLQs that have accurate full-width-at-half-maximum intensity of the broad component of the \hb~$\lambda4861$~emission line (hereafter, FWHM(\hb)), monochromatic luminosity at rest-frame 5100~\AA~(hereafter, \vLv), EW(\fetwo~$\lambda\lambda4434-4684$), and EW(\hb) measurements. Nine of these sources were obtained from SL15, seven from the GNIRS-DQS sample of Paper~I (see Section \ref{sec:sample}), and two from this work (see Appendix \ref{sec:spectroscopy}). SL15 compiled a sample of nine WLQs: SDSS J0836$+$1425, SDSS J1411$+$1402, SDSS J1417$+$0733, SDSS J1447$-$0203 \citep[][]{2010AJ....139..390P,Plotkin15}, SDSS J0945$+$1009 \citep[][]{2010MNRAS.404.2028H,Plotkin15}, SDSS J1141$+$0219, SDSS J1237$+$6301 \citep[][]{2009ApJ...699..782D,Shemmer10}, SDSS J1521$+$5202 \citep{2007ApJ...665.1004J, 2011ApJ...736...28W}, and PHL 1811 \citep{2007ApJS..173....1L}.

Table~\ref{table:WLQ} provides basic properties for the 18 WLQs in our sample. Column (1) provides the source name; Column (2) gives the systemic redshift determined from the peak of, in order of preference, the [\othree] $\lambda5007$, \magtwo~$\lambda2798$, and \hb~emission lines; Column (3) gives $\log \nu L_{\nu}$~($5100$~\AA); Column (4) gives FWHM(\hb); Column (5) gives \rfe~$\approx$~EW(\fetwo)/EW(\hb); Column (6) gives traditional \hb-based \mbh~estimates (following Equations~\ref{eq:mbh} and~\ref{eq:mbh_corrected}); Column (7) gives \fetwo-corrected \hb-based \mbhc~estimates (following Equations~\ref{eq:fetwocorrection} and~\ref{eq:mbh_corrected}); Column (8) gives traditional \hb-based \lledd~values (from Equation~\ref{eq:lledd}); Column (9) gives \fetwo-corrected \hb-based \lleddc~values (from Equation~\ref{eq:lledd}); Column (10) gives EW(\civ); Column (11) gives Blueshift(\civ); Columns (12) and (13) provide the references for the rest-frame optical and UV spectral measurements, respectively. All derived properties are discussed in detail in Section \ref{sec:equations}.

The two WLQs from \citet{Shemmer10}~and the two introduced in Appendix~\ref{sec:spectroscopy} do not have a reliable \civ~line measurement in the literature, hence we perform our own measurements from their SDSS spectra, following the procedure in \citet[hereafter Paper~II]{D23}. Briefly, we fit the \civ~emission line with a local, linear continuum and two independent Gaussians. These Gaussians are constrained such that the flux densities lie between 0 and twice the value of the peak of the emission line; the FWHM is restricted to not exceed 15000 \kms. Furthermore, we visually inspect the initial fit to correct for any additional residuals. The EW of the line emission can then be measured, as well as the blueshift, which is calculated from the difference between $\lambda1549$ and the rest-frame wavelength of the peak of the emission-line profile (see Equation~\ref{eq:blueshift}).

Our WLQs appear to possess stronger relative optical \fetwo~emission (indicated by the larger \rfe~values) compared to ordinary quasars from their respective samples. Since such sources are selected based only on their \civ~emission-line strength (EW(\civ) $< 10$~\AA), we are unable to assess any potential biases introduced by the rest-frame optical emission to their selection process. 

\subsection{Ordinary Quasar Sample Selection}\label{sec:sample}

In order to create a comprehensive comparison sample of quasars for our analysis, which requires measurements of both the \hb~and \civ~emission lines, we select two catalogs of ordinary quasars from the literature. For the high-redshift quasars ($1.5 \lesssim z \lesssim 3.5$), \civ~emission properties can be obtained from SDSS, but the \hb~emission line lies outside of the SDSS range, and therefore it has to be measured with NIR spectroscopy. In this redshift range, we utilize the GNIRS-DQS catalog in Paper~I. GNIRS-DQS is the largest and most comprehensive inventory of rest-frame optical properties for luminous quasars, notably the \hb, [\othree], and \fetwo~emission lines. To complement this sample of high-redshift, high-luminosity quasars, we include an archival sample of quasars in the low-redshift regime from the BL04 subsample also utilized in SL15. In this redshift range ($z< 0.5$), the \hb~emission properties can be obtained from optical spectra, but the \civ~emission-line properties are more difficult to obtain, and are available primarily from the {\em Hubble Space Telescope} (HST) and the {\em International Ultraviolet Explorer} (IUE) archives. Below, we briefly discuss the selection process for our ordinary quasar sample.

The GNIRS-DQS sources were selected to lie in three narrow redshift intervals, $1.55 \lesssim z \lesssim 1.65$, \hbox{$2.10 \lesssim z \lesssim 2.40$}, and $3.20 \lesssim z \lesssim 3.50$ to center the \hb$+$[\othree] spectral complex in the NIR bands covered by GNIRS (i.e., the $J$, $H$, and $K$ bands, respectively). In total, the survey comprises 260 sources with high-quality NIR spectra and comprehensive \hb, [\othree], and \fetwo~spectral measurements \citep[see][and Paper~I for more details]{Matthews21}. We exclude 64 BAL quasars, 16 radio-loud quasars (RLQs), and one quasar, SDSS~J114705.24$+$083900.6 that is both BAL and radio loud. We define RLQs as sources having radio-loudness values of $R > 100$ \citep[where $R$ is the ratio between the flux densities at 5~GHz and 4400~\AA;][]{1989AJ.....98.1195K}. RLQs and BAL quasars are excluded to minimize the potential effects of continuum boosting from a jet \citep[e.g.,][]{2014A&A...568A.114M} and absorption biases (e.g., see BL04), respectively. Two quasars, SDSS J073132.18$+$461347.0 and SDSS J141617.38$+$264906.1, are excluded due to a lack of \civ~measurements from Paper~II. In total, 177 GNIRS-DQS quasars are included in our analysis; of these, seven sources with \hbox{EW(\civ) $<10$~\AA} can be formally classified as WLQs (see Section \ref{sec:WLQs}). We adopt values of FWHM(\hb), \vLv,  EW(\hb), and EW(\fetwo) values from Paper~I. The latter two parameters are used to derive \rfe. Paper~II reports the EW(\civ) values and the wavelengths of the \civ~emission-line peaks for the quasars in Paper~I, which are then used to derive the Blueshift(\civ) values (see Section~\ref{sec:zsys}).

Sixty quasars at $z < 0.5$ from BL04 are added to our analysis from the 63 BL04 quasars in SL15. PG 0049$+$171, PG 1427$+$480, and PG 1415$+$451 are excluded due to a lack of published \fetwo~spectral measurements. The UV data in the BL04 sample comes, roughly equally, from both the {\sl HST} and the {\sl IUE} archives \citep[see,][]{2005MNRAS.356.1029B}. Throughout this work, we check whether including only {\sl HST} or {\sl IUE} data changes the conclusion of the paper, but we find no statistical difference in the results of Section~\ref{sec:results}. Therefore, we include both subsets in this work.  We obtain the FWHM(\hb), \vLv, and \rfe~values for the BL04 sources from \citet{BG92}, and their EW(\civ) and Blueshift(\civ) values from \citet{2005MNRAS.356.1029B}. The line measurements are expected to be roughly consistent across the different samples utilized in this work, since they all employed similar standard fitting procedures. Table~\ref{table:ordinaryq}~lists the basic properties of the ordinary quasars in our sample with the same formatting as Table~\ref{table:WLQ}.

\subsection{Systemic redshifts and the Blueshift(\civ)} \label{sec:zsys}

We derive the Blueshift(\civ) values of GNIRS-DQS sources from the observed wavelengths of the \civ~emission-line peaks reported in Paper~II and the systemic redshifts reported in Paper~I. The Blueshift(\civ) values are derived following Equation (2) in \citet{Dix20}
\begin{equation}
    \label{eq:blueshift}
    \frac{\Delta v}{\mathrm{km~s^{-1}}}= \left[\frac{c}{\mathrm{km~s^{-1}}}\right]\left(\frac{z_{\rm meas} - z_{\rm sys}}{1+z_{\rm sys}}\right),
\end{equation}
where $z_{\rm meas}$~is the redshift measured from the wavelength of the \civ~emission-line peak, and $z_{\rm sys}$ is the systemic redshift with respect to the [\othree], the \magtwo, or the \hb ~emission lines reported in Paper~I. In this work, we report the Blueshift(\civ) $\equiv -\Delta v$~values.

A non-negligible fraction ($\sim 1/3$) of luminous quasars  have extremely weak or undetectable [\othree] emission \citep[e.g.,][]{Netzer04}, so we must use alternative emission lines as the reference for $z_{\rm sys}$ (as was done for many ordinary GNIRS-DQS sources; see, Paper~I). In spite of the larger intrinsic uncertainties
associated with using the \magtwo~and \hb~emission lines as $z_{\rm sys}$ indicators \citep[$\sim 200$~km~s$^{-1}$~and $\sim 400$~km~s$^{-1}$, respectively;][]{2016ApJ...831....7S}, these uncertainties are typically much smaller than the Blueshift(\civ) values observed in the majority of luminous high-redshift quasars (see, Paper~I). Therefore, the lack of [\othree]-based $z_{\rm sys}$ values for such sources should not affect the conclusions of this work significantly. 

\subsection{\mbh~and \lledd~Estimates} \label{sec:equations}

Traditional estimation of single-epoch \mbh~values has made use of the reverberation-mapping (RM) scaling relationship between the size of the \hb-emitting region ($R_{\mathrm{H}\beta}$) and \vLv~\citep[e.g.,][]{1998ApJ...505L..83L,1999ApJ...526..579W,2005ApJ...629...61K,2013ApJ...767..149B}. In this work, we use the empirical scaling relation established by \citet{2013ApJ...767..149B} for consistency with other recent studies \citep[e.g.,][Paper~II]{2022MNRAS.tmp.1709M}:
\begin{equation}
    \label{eq:mbh}
    \textup{log}\left[\frac{R_{\rm H\beta}}{\textup{lt}-\textup{days}}\right] = (1.527\pm0.031) + (0.533\pm0.035)~\textup{log}~\ell_{44}
\end{equation}
where $\ell_{44} \equiv \nu L_{\nu} (5100$~\AA)~$/10^{44}$ erg~s$^{-1}$.

\begin{rotatetable*}
\begin{deluxetable*}{lccccccccccccc}
\label{table:WLQ}
\tablecolumns{13}
\tabletypesize{\scriptsize}
\tablecaption{Basic Properties of the WLQ Sample}
\tablewidth{0pt}
\tablehead{
\colhead{Quasar} &
\colhead{$z_{\rm sys}$} &
\colhead{$\log \nu L_{\nu}(5100$\,\AA$)$} &
\colhead{FWHM(\hb)} &
\colhead{\rfe} &
\colhead{$\log$~\mbh} &
\colhead{$\log$~\mbhc} &
\colhead{\lledd} &
\colhead{\lleddc} &
\colhead{EW(\civ)} &
\colhead{Blueshift(\civ)} &
\colhead{Optical} &
\colhead{\civ} \cr
\colhead{} &
\colhead{} &
\colhead{(erg s$^{-1}$)} &
\colhead{(\kms)} &
\colhead{} &
\colhead{(\msun)} &
\colhead{(\msun)} &
\colhead{} &
\colhead{} &
\colhead{(\AA)} &
\colhead{(\kms)} &
\colhead{Ref.\tablenotemark{a}} &
\colhead{Ref.\tablenotemark{a}} \cr
\colhead{(1)} &
\colhead{(2)} &
\colhead{(3)} &
\colhead{(4)} &
\colhead{(5)} &
\colhead{(6)} &
\colhead{(7)} &
\colhead{(8)} &
\colhead{(9)} &
\colhead{(10)} &
\colhead{(11)} &
\colhead{(12)} &
\colhead{(13)}
}
\startdata
SDSS~J010643.23$-$031536.4 & 2.248 &     46.51 &     6782 &  0.58 &       9.99 & 9.71 &         0.20 &   0.39 &     $7.6^{+0.6}_{-0.9}$ &           $1451^{+119}_{-60}$ &    1 &    2 \\
SDSS~J013136.44+130331.0 & 1.599 &     46.45 &     2294 &  0.78 &       9.02 & 8.67 &         1.63 &   3.67 &     $2.8^{+1.4}_{-2.0}$ &           $2320^{+819}_{-521}$ &    1 &    2 \\
SDSS~J013417.81$-$005036.2 & 2.270 &     46.45 &     5211 &  0.98 &       9.73 & 9.31 &         0.32 &   0.84 &     $7.3^{+0.7}_{-1.0}$ &           $2233^{+651}_{-414}$ &    1 &    2 \\
SDSS~J075115.43+505439.1 & 2.311 &     46.59 &     3077 &  3.05 &       9.35 & 8.19 &         1.05 &  15.04 &     $6.6^{+0.6}_{-1.0}$ &           $5953^{+234}_{-117}$ &    1 &    2 \\
SDSS~J083650.86+142539.0 & 1.749 &     45.93 &     2880 &  2.48 &       8.94 & 8.04 &         0.62 &   4.95 &     $4.2^{+0.3}_{-0.5}$ &           $2266\pm191$ &    3 &    3 \\
SDSS~J085337.36+121800.3 & 2.197 &     46.56 &     4502 &  0.28 &       9.66 & 9.48 &         0.47 &   0.73 & $7.7^{+1.1}_{-1.7}$ & $1166^{+363}_{-242}$ & 1 & 2 \\
SDSS~J085344.17+354104.5 & 2.183 &     46.40 &     4168 &  0.72 &       9.51 & 9.18 &         0.47 &   1.00 &     $4.3^{+0.8}_{-1.2}$ & $2053^{+1580}_{-1094}$& 1 & 2 \\
SDSS~J094533.98+100950.1 & 1.683 &     46.17 &     4278 &  2.00 &       9.41 & 8.66 &         0.35 &   2.03 &     $2.9^{+0.3}_{-0.6}$ & $5485\pm380$ & 3 & 3 \\
SDSS~J094602.31+274407.0 & 2.488 &     46.75 &     3833 &  1.65 &       9.63 & 8.94 &         0.79 &   3.82 &     $5.9^{+0.4}_{-0.6}$ & $9062^{+16}_{-11}$ & 1 & 2 \\
SDSS~J113747.64+391941.5 & 2.428 &     45.81 &     2518 &  3.31 &       8.76 & 7.57 &         0.72 &  10.99 &     $8^{+6}_{-9}$ & $3089^{+2050}_{-1236}$ & 4 & 4 \\
SDSS~J114153.33+021924.4 & 3.550 &     46.55 &     5900 &  3.25 &       9.89 & 8.67 &         0.27 &   4.60 &     $0.4^{+2}_{-4}$ & $-577^{+2461}_{-1484}$ & 5 & 6,4\\
SDSS~J123743.07+630144.7 & 3.490 &     46.35 &     5200 &  2.86 &       9.68 & 8.61 &         0.29 &   3.39 &     $1\pm2$ & $-970^{+1349}_{-845}$ & 5 & 4\\
SDSS~J141141.96+140233.9 & 1.754 &     45.64 &     3966 &  1.41 &       9.06 & 8.56 &         0.24 &   0.78 &     $3.8^{+0.8}_{-0.2}$ & $3142^{+370}_{-208}$ & 1 & 2 \\
SDSS~J141730.92+073320.7 & 1.716 &     45.91 &     2784 &  1.65 &       8.90 & 8.29 &         0.65 &   2.64 &    $2.5^{+2.1}_{-0.7}$ & $5321^{+4178}_{-872}$ & 1 & 2 \\
SDSS~J144741.76$–$020339.1 & 1.430 &     45.56 &     1923 &  1.60 &       8.39 & 7.83 &         0.96 &   3.52 &     $7.7^{+0.2}_{-1.3}$ & $1319^{+759}_{-381}$ & 1 & 2 \\
SDSS~J152156.48+520238.5 & 2.190 &     47.14 &     5750 &  1.64 &      10.19 & 9.48 &         0.52 &   2.69 &     $9.1\pm0.6$ & $4900\pm300$ \tablenotemark{b} & 7 & 7\\
SDSS~J213742.25$-$003912.7 & 2.294 &     45.75 &     2630 &  2.45 &       8.77 & 7.89 &         0.62 &   4.68 &     $3^{+1}_{-2}$ & $4986^{+867}_{-535}$ & 4 & 4 \\
PHL 1811 & 0.192 & 45.56 & 1943 & 1.29 \tablenotemark{c} & 8.40 & 7.94 & 0.94 & 2.70 & 6.6 & $1400\pm250$ & 8 & 8
\enddata
\tablenotetext{a}{Source of rest-frame optical--UV data, Column (12): $z_{\rm sys}$, $\nu L_{\nu}(5100$\,\AA$)$, FWHM(\hb), \rfe; Column (13): EW(\civ), and Blueshift(\civ). (1) Paper~I; (2) Paper~II; (3) \citet{Plotkin15}; (4) this work; (5) \citet{Shemmer10}; (6) \citet{2011ApJS..194...45S}; (7) \citet{2011ApJ...736...28W};(8) \citet{2007ApJS..173....1L}.}
\tablenotetext{b}{\citet{2011ApJ...736...28W}~also reported \hb-based Blueshift(\civ) $= 9400$ km s$^{-1}$. Here, we have opted to use a \magtwo-based value of Blueshift(\civ).}
\tablenotetext{c}{\citet{2007ApJS..173....1L}~reported the \rfe~value as being in the range 1.22$-$1.35. We have adopted a mean value of 1.29 for this work.}
\end{deluxetable*}
\end{rotatetable*}

\begin{rotatetable*}
\begin{deluxetable*}{lccccccccccccc} 
\label{table:ordinaryq}
\tablecolumns{13}
\tabletypesize{\scriptsize}
\tablecaption{Basic Properties of the Ordinary Quasar Sample}
\tablewidth{0pt}
\tablehead{
\colhead{Quasar} &
\colhead{$z_{\rm sys}$} &
\colhead{$\log \nu L_{\nu}(5100$\,\AA$)$} &
\colhead{FWHM(\hb)} &
\colhead{\rfe} &
\colhead{$\log$~\mbh} &
\colhead{$\log$~\mbhc} &
\colhead{\lledd} &
\colhead{\lleddc} &
\colhead{EW(\civ)} &
\colhead{Blueshift(\civ)} &
\colhead{Optical} &
\colhead{\civ} \cr
\colhead{} &
\colhead{} &
\colhead{(erg s$^{-1}$)} &
\colhead{(\kms)} &
\colhead{} &
\colhead{(\msun)} &
\colhead{(\msun)} &
\colhead{} &
\colhead{} &
\colhead{(\AA)} &
\colhead{(\kms)} &
\colhead{Ref.\tablenotemark{a}} &
\colhead{Ref.\tablenotemark{a}} \cr
\colhead{(1)} &
\colhead{(2)} &
\colhead{(3)} &
\colhead{(4)} &
\colhead{(5)} &
\colhead{(6)} &
\colhead{(7)} &
\colhead{(8)} &
\colhead{(9)} &
\colhead{(10)} &
\colhead{(11)} &
\colhead{(12)} &
\colhead{(13)}
}
\startdata
PG 0003+199 & 0.026 &     44.07 &     1640 &  0.62 &       7.46 &  7.36 &         0.33 &   0.41 &                  $60.1$\tablenotemark{b} &                 $-102$ &    3 &    4 \\            
SDSS J001018.88+280932.5 & 1.613 &     46.27 &     3189 &  0.06 &       9.21 &  9.13 &         0.70 &   0.86 &    $61.0^{+0.5}_{-0.8}$ &      $203^{+22}_{-15}$ &    1 &    2 \\
SDSS J001453.20+091217.6 & 2.335 &     46.36 &     6428 &  0.72 &       9.87 &  9.54 &         0.19 &   0.40 &    $39.0^{+3.3}_{-5.0}$ &    $825^{+397}_{-266}$ &    1 &    2 \\
SDSS J001813.30+361058.6 & 2.333 &     46.46 &     4896 &  0.55 &       9.68 &  9.41 &         0.36 &   0.68 &    $25.8^{+1.1}_{-1.6}$ &   $2689^{+203}_{-136}$ &    1 &    2 \\
SDSS J001914.46+155555.9 & 2.267 &     46.34 &     4033 &  0.17 &       9.45 &  9.32 &         0.47 &   0.64 &    $44.5^{+0.9}_{-1.3}$ &     $372^{+110}_{-74}$ &    1 &    2 \\
PG 0026+129 & 0.145 &     45.13 &     1860 &  0.51 &       8.13 &  7.98 &         0.68 &   0.95 &                  $19.3$ &                 $-120$ &    3 &    4 \\
SDSS J002634.46+274015.5 & 2.247 &     46.38 &     4420 &  0.00 &       9.55 &  9.48 &         0.41 &   0.49 & $134.6^{+10.1}_{-15.0}$ &    $400^{+416}_{-279}$ &    1 &    2 \\
SDSS J003001.11-015743.5 & 1.590 &     46.10 &     4028 &  0.26 &       9.32 &  9.18 &         0.37 &   0.52 &    $52.7^{+1.9}_{-2.8}$ &    $1279^{+139}_{-93}$ &    1 &    2 \\
SDSS J003416.61+002241.1 & 1.631 &     46.24 &     5527 &  0.44 &       9.67 &  9.46 &         0.23 &   0.38 &    $28.5^{+0.3}_{-0.5}$ &      $597^{+86}_{-58}$ &    1 &    2 \\
SDSS J003853.15+333044.3 & 2.373 &     46.39 &     4297 &  0.50 &       9.53 &  9.28 &         0.44 &   0.78 &    $13.8^{+1.0}_{-1.5}$ &    $670^{+947}_{-635}$ &    1 &    2 
\enddata
\tablenotetext{a}{Source of rest-frame optical--UV data, Column (12): $z_{\rm sys}$, $\nu L_{\nu}(5100$\,\AA$)$, FWHM(\hb), \rfe~; Column (13): EW(\civ), and Blueshift(\civ). (1) Paper~I; (2) Paper~II; (3) \citet{BG92}; (4) \citet{2005MNRAS.356.1029B}.}
\tablenotetext{b}{There are no errors reported for EW(\civ) and Blueshift(\civ) values for PG quasars in \citet{2005MNRAS.356.1029B}.}
\tablecomments{Only the first ten lines are shown; the entire table is available in the electronic version.}
\end{deluxetable*}
\end{rotatetable*}

However, the \hb~RM sample was subsequently determined to be biased toward objects with strong, narrow [\othree] emission-lines, and, in effect, is biased in favor of low-accretion-rate broad-line AGNs \citep[see, e.g., ][]{1994ASPC...69..147R,2014Natur.513..210S}. Recent RM campaigns aimed at minimizing such bias, such as the Super-Eddington Accreting Massive Black Hole (SEAMBH; \citeauthor{2014ApJ...782...45D}~\citeyear{2014ApJ...782...45D,2016ApJ...818L..14D,2018ApJ...856....6D}) and the SDSS-RM project \citep{2015ApJS..216....4S}, found deviations from the traditional size-luminosity relationship. In particular, SEAMBH found a population of rapidly accreting AGNs with a BELR size up to 3-8 times smaller than predicted by Equation~\ref{eq:mbh}, which implies an overestimation of super-Eddington-accreting \mbh~values from single-epoch spectra by the same factor. We apply a \rfe~correction to the traditional \hb-based \mbh~estimation, a method developed by \citet{2019ApJ...886...42D}. The \rfe~paramater has been shown to correlate with \lledd~\citep[e.g.,][]{2007ApJ...654..754N}.

For the \fetwo-corrected values of \mbh~(hereafter, \mbhc), we apply the size-luminosity scaling relation for $R_{\mathrm{H}\beta}$ following Equation (5) of \citet{2019ApJ...886...42D}:
\begin{equation}
    \label{eq:fetwocorrection}
    \begin{aligned}
        &\textup{log}\left[\frac{R_{\rm H\beta,~corr}}{\textup{lt$-$days}}\right] \\
        & = (1.65\pm0.06) + (0.45\pm0.03)~\textup{log}~\ell_{44} \\
    & + (-0.35\pm0.08)~R_{\rm Fe~II}.
    \end{aligned}
\end{equation}

Subsequently, \mbh~(\mbhc) can be estimated using the following relationship:
\begin{equation} \label{eq:mbh_corrected}
    \begin{aligned}
        &\frac{M_{\rm BH}~(M_{\rm BH,~corr})}{M_{\odot}} \\
        & = f \left[\frac{R_{\rm BELR}}{\rm pc}\right] \left[\frac{\Delta V}{\rm km~s^{-1}}\right]^2 \left[\frac{G}{\rm pc~M_{\odot}^{-1}~(km~s^{-1})^2}\right]^{-1} \\
        & \approx 1.5 \left[\frac{R_{\rm H\beta}~(R_{\rm H\beta,~corr})}{\rm pc}\right] \left[\frac{\rm FWHM(H\beta)}{\rm km~s^{-1}}\right]^2 \\ & \cdot \left[\frac{4.3 \times 10^{-3}}{\rm pc~M_{\odot}^{-1}~(km~s^{-1})^2}\right]^{-1},
    \end{aligned}
\end{equation}
where we adopt $f=1.5$ for the virial coefficient, consistent with results from \citet{2014ApJ...789...17H,2019MNRAS.488.1519Y,2020ApJ...901..133Y,2022MNRAS.tmp.1709M}, $R_{\rm BELR} \approx R_{\rm H\beta}$ ($R_{\rm H\beta,~corr}$) is the size-luminosity relation from Equation~\ref{eq:mbh}~(\ref{eq:fetwocorrection}), $\Delta V$ is the velocity width of the emission line, which is taken here as FWHM(\hb), assuming Doppler broadening \citep{1999ApJ...526..579W}, and $G$ is the gravitational constant. 

The \lledd~parameter can be computed from the corresponding \mbh~value following Equation (2) of \citet{Shemmer10} assuming that $L_{\rm Edd}$~is computed for the case of solar metallicity:

\begin{equation}
\label{eq:lledd}
    \begin{aligned}
        & L/L_{\rm Edd}~(L/L_{\rm Edd,~corr})\\
        &=1.06 f(L) \left[\frac{\nu
    L_{\nu}(5100\,\mbox{\AA})}{10^{44}\,{\rm
      ergs\,s^{-1}}}\right] \left[\frac{M_{\rm BH}~(M_{\rm BH,~corr})}{10^6 M_{\sun}}\right]^{-1},        
    \end{aligned}
\end{equation}
where $f(L)$ is the luminosity-dependent bolometric correction to \vLv, derived from Equation (21) of \citet{Marconi04}. 

We note that a wide range of bolometric corrections for quasars is available in the literature \citep[e.g.,][]{2006ApJS..166..470R,2010MNRAS.408.1598N,2012MNRAS.422..478R,2019MNRAS.488.5185N}. However, in general, the range of these corrections is not large enough to affect the conclusion of our work. For example, \citet{2022MNRAS.tmp.1709M} recently used a constant bolometric correction of $L_{\rm Bol} /$\vLv~$\sim 9$; the bolometric corrections we derive are in the range of $\sim 5$-$6$, which results in a relatively small {\em systematic} offset in the derivation of the \lledd~parameter.

The uncertainties associated with the corrected \mbh~and \lledd~values in this work are estimated to be at least $\sim 0.3$~dex \citep[see Table 2 of][]{2022MNRAS.tmp.1709M}, but could be much larger ($\sim 0.4-0.6$~dex) for high \lledd~objects such as WLQs (see also, SL15). 

\section{Results and Discussion}\label{sec:results}

\begin{figure*}[t!]
    \centering
    \includegraphics[width=0.49\linewidth]{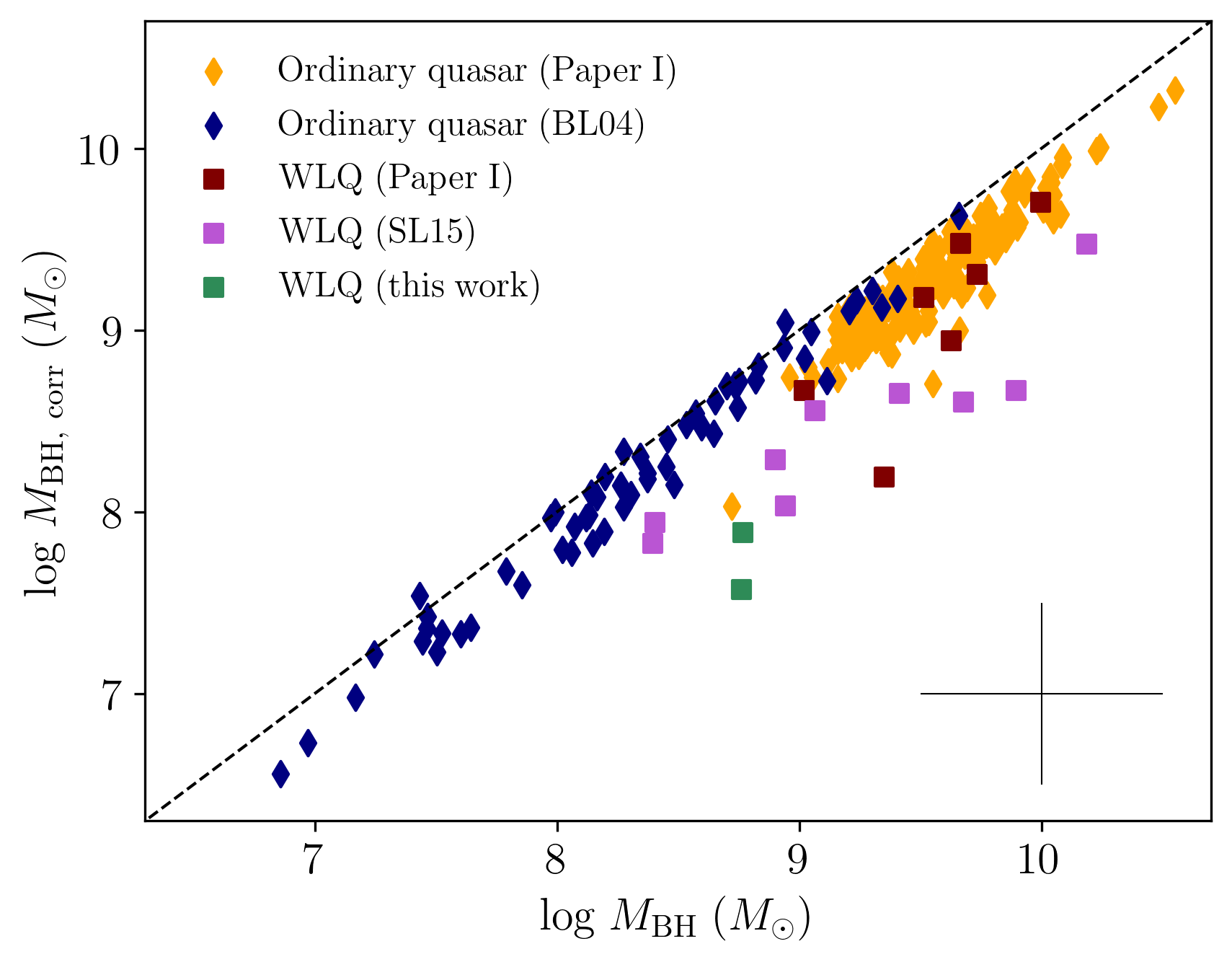}
    \includegraphics[width=0.49\linewidth]{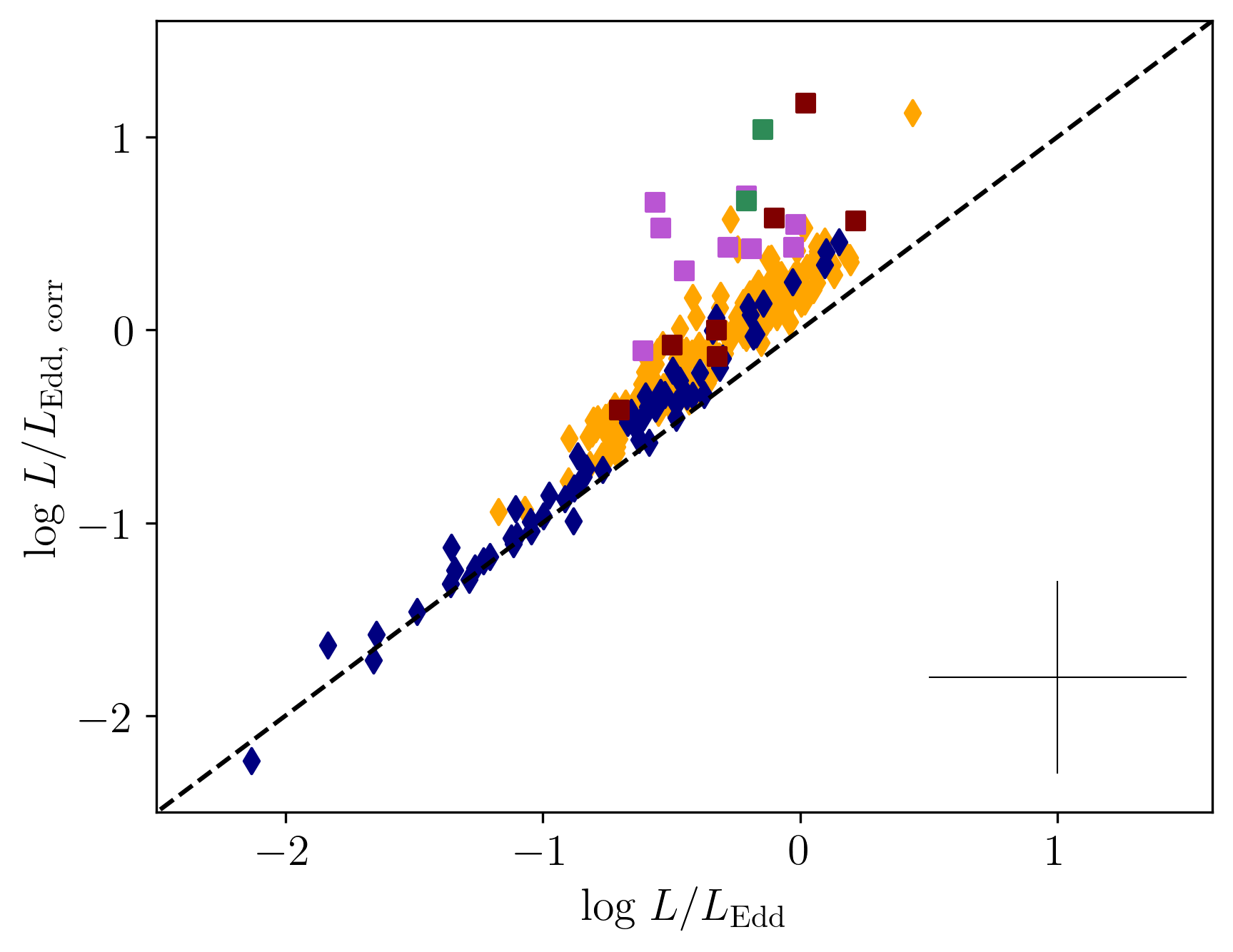}
    \caption{Black-hole mass (left panel) and accretion rate (right panel) calculated using the traditional (x-axis) and \rfe-corrected (y-axis) BELR size-luminosity relation for all quasars in our analysis. Diamonds mark ordinary quasars and squares mark WLQs. The dashed lines represent a one-to-one relation between the two methods. Typical uncertainties of 0.5 dex on the \mbh~and \lledd~values are displayed in the bottom right corner of each panel. The traditional relation overestimates \mbh~in rapidly-accreting quasars by up to an order of magnitude. In turn, the traditional relation underestimates \lledd~by a similar factor. In particular, the \rfe-corrected accretion rates are much larger for a considerably larger fraction of sources in the WLQ subset than in the ordinary quasars, due to their larger \rfe~values.}
    \label{fig:canon_correct}
\end{figure*}

\subsection{Black Hole Masses and Accretion Rates}\label{sec:characteristics}

For the 248 quasars included in this work, we determine the virial \hb-based \mbhc~and corresponding \lleddc~values from their derived optical properties, following the \fetwo-corrected BELR size-luminosity relation of Equation~\ref{eq:fetwocorrection}, applied to Equations~\ref{eq:mbh_corrected} and \ref{eq:lledd}. We also calculate these quasars' \mbh~and \lledd~values following the traditional BELR size-luminosity relation of Equation \ref{eq:mbh} to compare the two methods for estimating black-hole masses and accretion rates.

Figure~\ref{fig:canon_correct}~presents the traditional versus corrected \mbh~and \lledd~values for the quasars in our sample, following the procedure of \citet{2022MNRAS.tmp.1709M}. The \hb-based \mbhc~values of ordinary quasars show small systematic deviations from the traditional BELR size-luminosity relation estimates (less than a factor of two for 222 out of 230 quasars). On the other hand, for a majority of the WLQs, due to the relative weakness in \hb~emission compared to the \fetwo~emission, \mbhc~values deviate significantly from the traditional relation, by up to one order of magnitude. Since \lledd~is inversely proportional to \mbh, the  \lleddc~values are enhanced by a similar factor. This result is in line with the \citet{2022MNRAS.tmp.1709M} finding of a larger deviation from the one-to-one relation in high-accretion-rate quasars. 

\subsection{The Anti-correlation between EW(\civ) and \lledd}
\label{sec:MBE}

\begin{figure*}[t]
    \centering
    \includegraphics[width=0.9\linewidth]{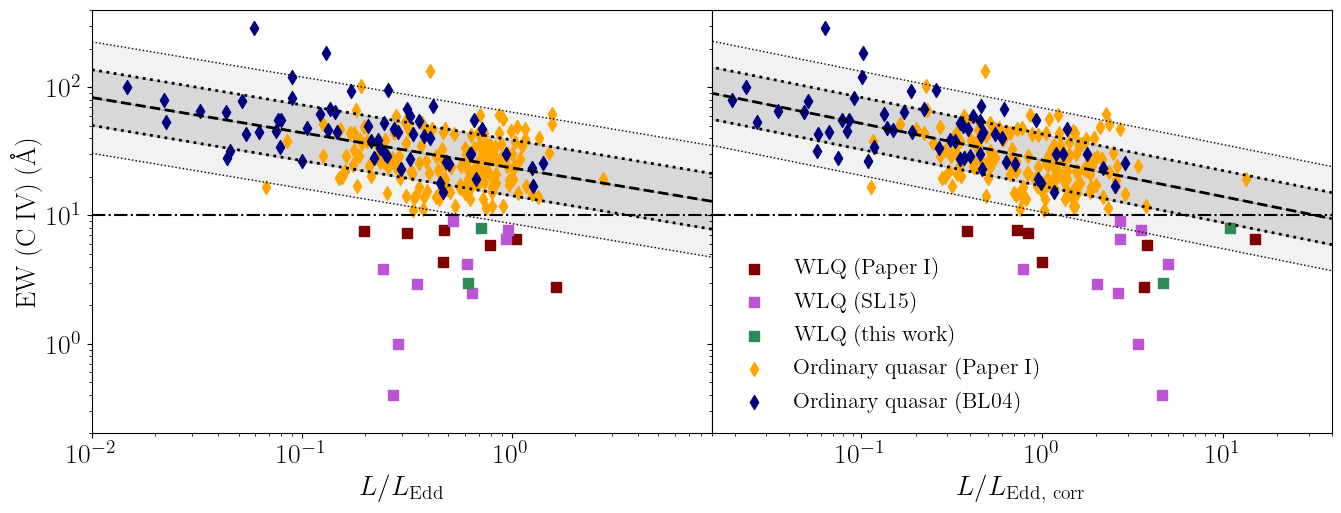}
    \caption {Correlation between EW(\civ) and \lledd~of ordinary quasars (diamonds) and WLQs from Table~\ref{table:WLQ} (squares). The left panel presents the traditional \lledd~values, and the right panel displays the  \fetwo-corrected \lleddc~values. The dotted-dashed lines represent the EW threshold below which objects are defined as WLQs. The correlation for the ordinary quasar sample, obtained by fitting a linear model, is shown as a dashed line. The shaded regions represent the 1- and 2-$\sigma$~deviations from the fitted correlation. Correcting the traditional \lledd~values results in a stronger anti-correlation expected by the MBE (see Table~\ref{tab:Spearman}); however, WLQs' \lleddc~values are still considerably (more than an order of magnitude) over-predicted by the MBE, suggesting that EW(\civ) is not the sole indicator of quasars' accretion rates.}
    \label{fig:baldwin}
\end{figure*}

We use our sample to explore the anti-correlation between EW(\civ) and \hb-based \lledd~previously studied in SL15 (i.e., the MBE), as well as with \lleddc. Figure~\ref{fig:baldwin} shows EW(\civ) plotted against the traditional \lledd~values (left) and against the \fetwo-corrected \lleddc~values (right). The first four rows of Table~\ref{tab:Spearman} present the respective Spearman-rank correlation coefficients ($r_{\rm S}$) and chance probabilities ($p$) of the ordinary quasar sample and the complete sample, including WLQs, for the correlation involving EW(\civ). We detect significant anti-correlations between EW(\civ) and \lledd~both with and without WLQs (i.e., $p \ll 1 \%$). However, the anti-correlation for the sample including WLQs is slightly weaker than without WLQs (both $p$ values are roughly similar, but $r_{\rm S}$ increases slightly). Our result reaffirms findings by SL15, who found WLQs to be outliers in this relation.

With a \fetwo~correction, the \lleddc~values provide a significantly stronger anti-correlation with EW(\civ) as the $r_{\rm S}$ value decreases from $-0.36$ (for the \lledd~case) to $-0.48$. Furthermore, the inclusion of WLQs no longer spoils the Spearman-rank correlation; in fact, the $p$~value remains extremely low ($p =4.02 \times 10^{-20}$ for the entire sample), and the $r_{\rm S}$ value decreases from $-0.48$ to $-0.54$, indicative of a stronger anti-correlation. Nevertheless, the \lleddc~values of most of the WLQs in our sample still appear considerably smaller than a linear model would suggest (see Figure~\ref{fig:baldwin}). To quantify the deviation of WLQs from the MBE, we fit a simple linear model, without considering the errors, to the $\log$~EW(\civ) and $\log$~\lleddc~values of the ordinary quasar sample. Our WLQs deviate from the best-fit model by a mean of $\sim 3.4 \sigma$, with a range in deviation from $1.08 \sigma$~to $8.02 \sigma$. Such a discrepancy paints WLQs as significant outliers in this correlation.

We also explore whether a bolometric luminosity correction based on the peculiarity of WLQs could account for this discrepancy. Although several methods for correcting bolometric luminosity are available in the literature \citep[e.g.,][]{2006ApJS..166..470R,2010MNRAS.408.1598N,2012MNRAS.422..478R,2019MNRAS.488.5185N}, if the Eddington ratios of WLQs were to be reliably predicted by the MBE, these corrections must be up to $\sim 10^5$~times larger than those of \citet{Marconi04} (as in the case of SDSS~J1141$+$0219 with EW(\civ) $= 0.4$~\AA). Such a discrepancy is larger than the difference expected by any of the current bolometric correction methods in the literature. These results reveal that EW(\civ) is likely not the sole indicator of accretion rate in all quasars, in agreement with SL15.

\subsection{The \cdist~as an Indicator of \lledd}
\label{sec:distance}

\citet{2020ApJ...899...96R} used an independent component analysis (ICA) technique to analyze the spectral properties of the \civ~emission line in 133 quasars from the SDSS-RM project \citep{2015ApJS..216....4S}. In particular, they fitted a piece-wise polynomial to trace the positions of these sources on the EW(\civ) and the Blueshift(\civ) plane. The projected position of a quasar along this curve is defined as its `\cdist'. 
To calculate the value of this \cdist~parameter, we follow the procedure summarized in R22 and detailed in \cite{CIVdist}. In short, we first transform the values of the two axes (EW(\civ) and Blueshift(\civ)) to lie between 0 and 1, using the \texttt{MinMaxScaler} function within \texttt{scikit-learn} \citep{scikit-learn}. Then, the \cdist~values are measured relative to the first point of the best-fit curve, located at \hbox{EW(\civ)~$\approx 316$~\AA} and \hbox{Blueshift(\civ)~$\approx 50$~km s$^{-1}$.}
This parameter essentially indicates the location along a non-linear first principal component of the \civ~parameter space, and encodes information about the physics of the \civ-emitting gas \citep[e.g.,][]{2011AJ....141..167R,2021AJ....162..270R,2019A&A...630A..94G}. 

\begin{figure*}[h]
    \includegraphics[width=0.5\linewidth]{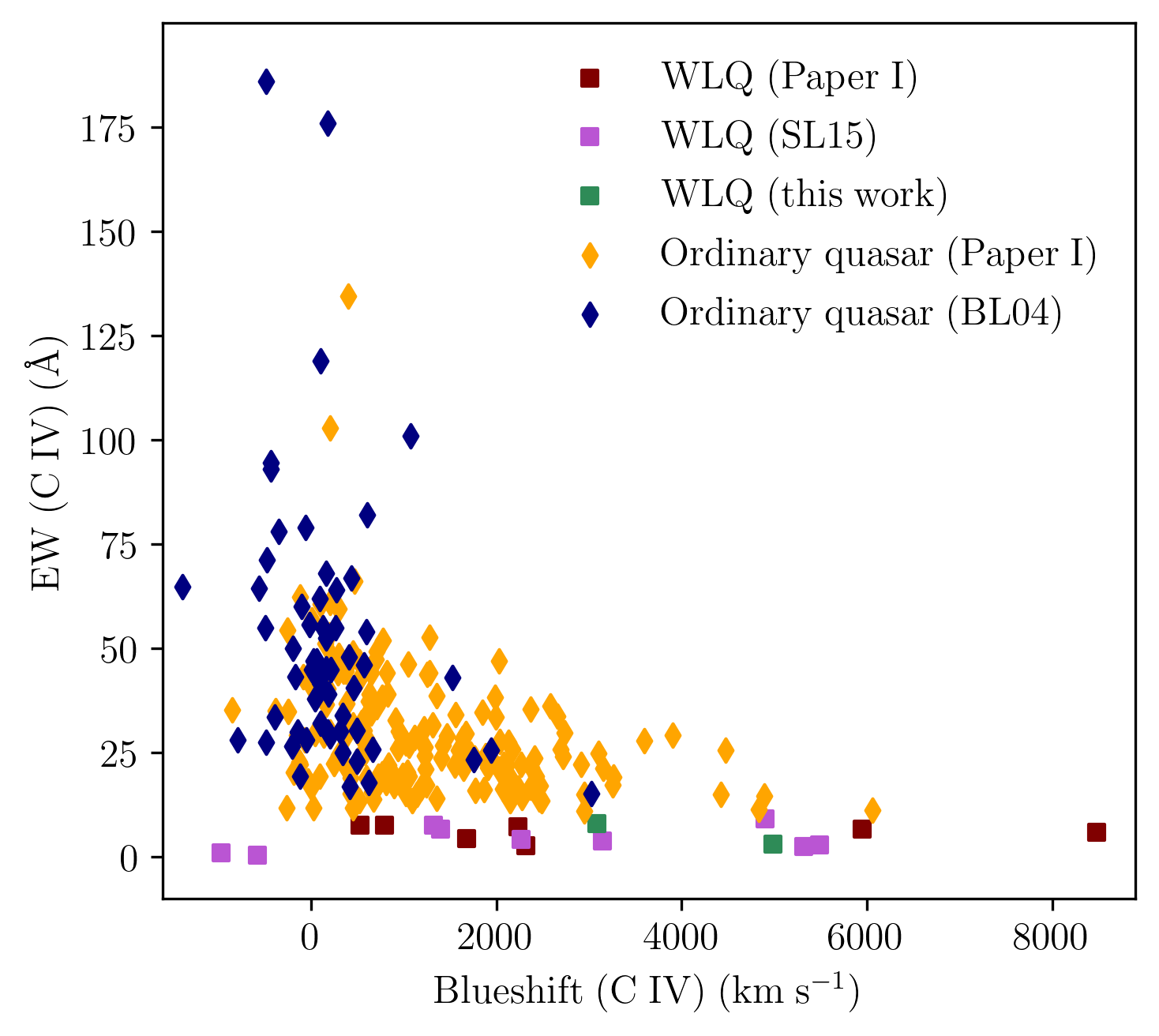}
    \includegraphics[width=0.49\linewidth]{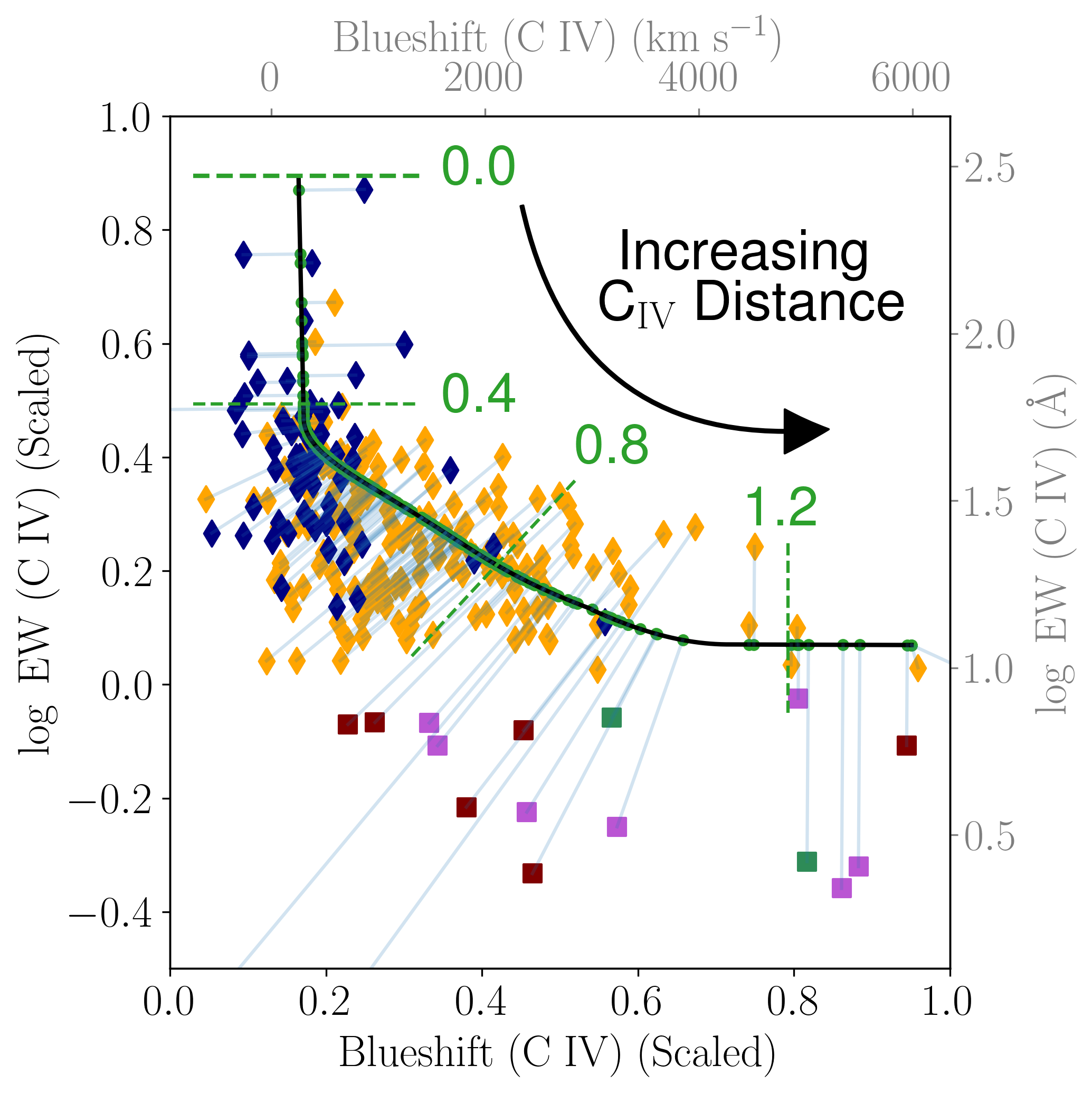}
    \caption{Left panel: distribution of EW(\civ) versus Blueshift(\civ) for our sample. One quasar from BL04, PG~1202$+$281, is not shown in the left panel, due to its extremely high EW(\civ) $= 290$~\AA. Right panel: illustration of the \cdist~parameter. The data are first scaled so that the two axes share the same limit, then each data point is projected onto the best-fit curve obtained from R22. The \cdist~value of each quasar is defined as its projected position (green point) along the solid black curve. Three of the WLQs, SDSS~J114153.33$+$021924.4, SDSS~J123743.07$+$630144.7, SDSS~J094602.31$+$274407.0, and one ordinary quasar, PG~1202$+$281, are not shown in the right panel, for clarity, but only their projected positions onto the curve are relevant to our results.}
    \label{fig:CIV_distance}
\end{figure*}

The left panel of Figure~\ref{fig:CIV_distance} shows the distribution of EW(\civ) versus Blueshift(\civ) of the 248 quasars in our sample. The right panel of Figure~\ref{fig:CIV_distance} shows the same distribution in scaled space, following the procedure in \citet{CIVdist}, and the piece-wise polynomial best-fit curve from Figure 2 of R22. Even though our sources are drawn from samples that are different from those of R22, the best-fit curve traces the C IV parameter space of sources across wide ranges of redshifts and luminosities. Since all quasars in our sample are selected photometrically, either in optical (for GNIRS-DQS quasars) or UV (for BL04 quasars) surveys, and were not selected based on their spectroscopic characteristics, there are no known biases associated with their selection, and thus they are expected to trace the \civ~parameter space in a similar manner to larger samples of quasars in other studies \citep[e.g., see also][]{2020MNRAS.492.4553R}. 

\begin{figure*}[h!]
    \centering
    \includegraphics[width=0.9\linewidth]{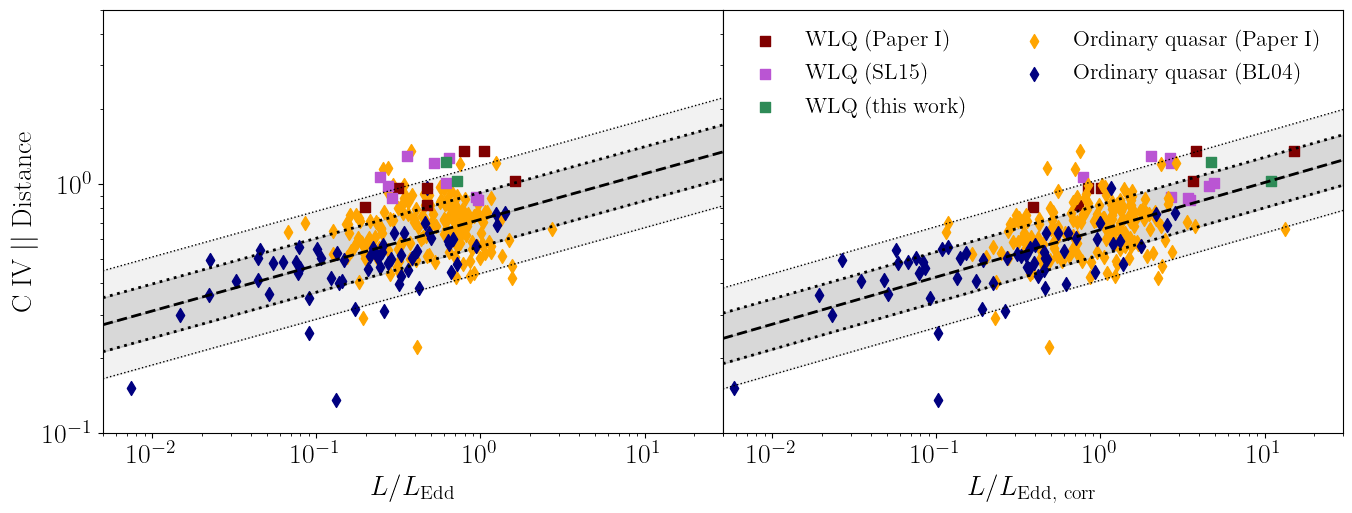}
    \caption{\cdist~versus \lledd~of 248 quasars in our sample. In the left panel, the \cdist~values are plotted against the traditional \hb-based \lledd~parameter, and in the right panel, against the \fetwo-corrected \hb-based \lleddc~parameter. The ordinary quasar PG~1202$+$281 with \lleddc~$= 0.06$~and \cdist~$= 0.02$ is not plotted, for clarity. The correlation for the ordinary quasar sample, obtained by fitting a linear model, is shown as a dashed line. The shaded regions represent the 1- and 2-$\sigma$~deviations from the fitted correlation. While using the traditional size-luminosity relation to estimate accretion rates already yields a strong correlation, the \fetwo-corrected accretion rates show a much stronger correlation with the \cdist~parameter for {\em all} quasars. Furthermore, this parameter serves as a better predictor for \lleddc~than for \lledd.}
    \label{fig:distance_corr}
\end{figure*}

While the EW(\civ) parameter, on its own, is not an ideal accretion-rate indicator, the \cdist~parameter appears to provide a robust indication of the accretion rates for all quasars including WLQs. We plot the \cdist~versus \hb-based \lledd~(left) and \lleddc~(right) for all quasars in our sample in Figure~\ref{fig:distance_corr}. The last four rows of Table~\ref{tab:Spearman} provide the Spearman-rank correlation coefficients and chance probabilities for the correlations involving the \cdist. Both the \lledd~and \lleddc~are significantly correlated with the \cdist~parameter (i.e., $p \ll 1$\%). 

In the case of \cdist~versus \lleddc, the correlation coefficient is considerably larger than the correlation involving \lledd~(0.57 versus 0.36), indicating the importance of the \fetwo~correction to \mbh. Furthermore, the inclusion of WLQs in the sample both strengthens the correlation ($r_{\rm S}$~increases from 0.52 to 0.57 while the $p$~value remains extremely small, $<10^{-16}$), and allows the high-\lleddc~end of the correlation to be more populated. There is also no significant deviation of the WLQs from this correlation, as opposed to their behavior in the MBE (see, Figure~\ref{fig:baldwin}) as well as in the \cdist~versus traditional \lledd~(see left panel of Figure~\ref{fig:distance_corr}). To quantify this effect, we fit a linear model to the \cdist~and \lledd~(\lleddc) space, taking into account only the ordinary quasars. Then, we calculate the mean scatter of the WLQs from this line. In the case of \lledd, we find the deviation from the best-fit line to range from $0.62 \sigma$~to $2.96 \sigma$, and the mean deviation to be $\sim 1.8 \sigma$. Meanwhile, the deviation in the case of \lleddc~ranges from $0.01 \sigma$~to $2.18 \sigma$, with a mean deviation of $\sim 1.1 \sigma$. Thus, using \lleddc~not only results in a stronger correlation with \cdist, but \cdist~also serves as a better predictor for \lleddc~than for \lledd.

The right panel of Figure~\ref{fig:distance_corr} shows that WLQs are not a disjoint subset of quasars in the UV$-$optical space \citep[see also,][]{2018A&A...618A.179M}. Our results indicate that WLQs possess relatively high accretion rates, due not only to their extremely weak \civ~lines, but rather to their relatively large values of the \cdist~parameter. Similarly, we observe quasars with high accretion rates (and large values of \cdist) that do not necessarily possess extremely weak \civ~lines, some of which have Eddington ratios that are larger than those of several WLQs. Finally, while we are unaware of a large population of quasars that deviate significantly from the correlations of Figure~\ref{fig:distance_corr}, a future examination of, e.g., \hb-based \lledd~values of quasars with very large EW(\civ) \citep[e.g.,][]{2022ApJ...934...97F} is warranted to further test our results.

In this work, we show that the \civ~and \hb~parameter space provides important diagnostics for quasar physics. In particular, we found that the \cdist~can serve as a robust predictor of quasar's accretion rate, especially after a correction based on \rfe~is applied. Within the limits of our sample, we also find that WLQs are not a disjoint subset of the Type~1 quasar population, but instead lie preferentially toward the extreme end of the \civ-\hb~parameter space. 

\begin{deluxetable*}{llccc}
\tablecolumns{5}
\tablecaption{Spearman-Rank Correlation Coefficients}
\tablehead{
\colhead{Correlation} &
\colhead{Sample} &
\colhead{$N$} &
\colhead{$r_{\rm S}$} &
\colhead{$p$}
}
\startdata
EW(\civ)-\lledd & Ordinary & 230 &  $-0.36 $ & $ 1.39 \times 10^{-8}$ \\
EW(\civ)-\lledd & All & 248 & $-0.35$ & $1.10 \times 10^{-8}$ \\
EW(\civ)-\lleddc & Ordinary & 230 &  $-0.48 $ & $ 6.82 \times 10^{-15}$ \\
EW(\civ)-\lleddc & All & 248 & $-0.54$ & $4.02 \times 10^{-20}$ \\
\cdist-\lledd & Ordinary & 230 & $0.37$ & $6.50 \times 10^{-9}$ \\
\cdist-\lledd & All & 248 & $0.36$ & $4.17 \times 10^{-9}$ \\
\cdist-\lleddc & Ordinary & 230 & $0.52$ & $3.16 \times 10^{-17}$ \\
\cdist-\lleddc & All & 248 & $0.57$ & $1.97 \times 10^{-22}$
\enddata
\tablecomments{The last three columns represent the number of sources in each correlation, the Spearman-rank correlation coefficient, and the chance probability, respectively.}
\label{tab:Spearman}
\end{deluxetable*}

\section{Conclusions}\label{sec:conclusions}

We compile a statistically meaningful sample of ordinary quasars and WLQs to study the dependence of quasar accretion rates, corrected for the relative strength of \fetwo~emission with respect to \hb, upon source location in the \civ~parameter space. Utilizing 18 WLQs, 16 of which are obtained from the literature and two of which are presented in this work, we confirm the findings of \citet{2022MNRAS.tmp.1709M} that the traditional approach to estimating the Eddington ratio for rapidly-accreting quasars systematically underestimates this property by up to an order of magnitude compared to \fetwo-corrected values of this parameter.

Using the \fetwo-corrected values of \hb-based \lledd, we investigate the correlation between this parameter and the \civ~parameter space. We confirm and strengthen the SL15 results by finding that WLQs spoil the anti-correlation between EW(\civ) and \hb-based \lledd~for quasars, whether the latter parameter is estimated using the traditional method, or whether a correction based on \fetwo~emission is employed in the \mbh~ estimate. In keeping with SL15, we conclude that the EW(\civ) cannot be the sole indicator of accretion rate in quasars.

We also investigate the relationships between a recently-introduced parameter, the \cdist, which is a combination of EW(\civ) and Blueshift(\civ), and the traditional \hb-based \lledd~and the \fetwo-corrected \lleddc. Such relationships yield strong correlations, especially in the case of \fetwo-corrected \lleddc, and can accommodate {\em all} the quasars in our sample. Our finding suggests that WLQs are not a disjoint subset of sources from the general population of quasars. We find that many WLQs have extremely high accretion rates which is indicated by their preferentially higher values of the \cdist~parameter. Similarly, we find several quasars in our sample that possess high Eddington ratios, and correspondingly large values of the \cdist, that do not have extremely weak \civ~lines; some of these sources display Eddington ratios that are larger than those of a subset of our WLQs.

In the context of the \civ~parameter space, it will be interesting to investigate whether the extreme \xray~properties of WLQs are the result of extremely large \cdist~values rather than resulting only from extremely weak \civ~lines. Such a test would require \xray~coverage of a large sample of sources with \hb+\fetwo~data across the widest possible \civ~parameter space such as the GNIRS-DQS sample of Paper~I. It would also be useful to determine whether the weakness of the broad \lya$+$\nv~emission line complex (from which the first high-redshift WLQs were identified) also correlates with \cdist, which will require rest-frame ultraviolet spectroscopy \citep{2022ApJ...929...78P}. The results of these investigations will shed new light on the connection between the quasar accretion rate and the physics of the inner accretion disk and BELR.

\begin{acknowledgements} 
This work is supported by National Science Foundation (NSF) grants AST-1815281 (B.M.M., O.S., C.D.), AST-1815645 (M.S.B., A.D.M., J.M.). G.T.R. was supported in part by NASA through a grant (HST-AR-15048.001-A) from the Space Telescope Science Institute, which is operated by the Association of Universities for Research in Astronomy, Incorporated, under NASA contract NAS5-26555. W.N.B. acknowledges support from NSF grant AST-2106990. B.L. acknowledges financial support from the National Natural Science Foundation of China grant 11991053. B.T. acknowledges support from the European Research Council (ERC) under the European Union's Horizon 2020 research and innovation program (grant agreement 950533) and from the Israel Science Foundation (grant 1849/19). We thank an anonymous referee for the insightful comments and suggestions which improved the clarity of this paper. This research has made use of the NASA/IPAC Extragalactic Database (NED), which is operated by the Jet Propulsion Laboratory, California Institute of Technology, under contract with the National Aeronautics and Space Administration.
\end{acknowledgements}

\software{
IRAF \citep{1986SPIE..627..733T},
matplotlib \citep{matplotlib}, numpy \citep{numpy1,numpy2}, pandas \citep{pandas2}, scipy \citep{scipy}, scikit-learn \citep{scikit-learn}.
}

\bibliography{main}
\bibliographystyle{aasjournal}

\appendix
\twocolumngrid

\section{NIR Spectroscopy of SDSS~J1137$+$3919 and SDSS~J2137$-$0039}
\label{sec:spectroscopy}

\renewcommand\thefigure{\thesection.\arabic{figure}}
\renewcommand\thetable{\thesection.\arabic{table}}
\setcounter{figure}{0}
\setcounter{table}{0}

SDSS J113747.64$+$391941.5 and SDSS J213742.25$-$003912.7 (hereafter, SDSS J1137$+$3919 and SDSS J2137$-$0039, respectively) are two WLQs with redshifts suitable for observing the H$\beta$ line in the $H$-Band. Observations of these quasars were carried out by the \geminiN~Observatory using GNIRS throughout four observing runs between 2014 March 14 and 2014 August 4, under program GN-2014A-Q-47. The observation log appears in Table~\ref{table:obs}. For both targets, we used the Short Blue Camera, with spatial resolution 0.$''$15~pix$^{-1}$, and a $1.0''$ slit to achieve a spectral resolution of $R\sim 600$. An $H$-filter was applied, producing a spectral range of 1.5 - 1.8 $\mu$m, corresponding to rest-frame $\sim4500 - 5300$ \AA. Exposure times for each subintegration were 238~s and 220~s, and the total integration times were 7140~s and 7040~s for SDSS J1137$+$3919 and SDSS J2137$-$0039, respectively. These observations were performed using the standard ``ABBA" nodding pattern of the targets along the slit in order to obtain primary background subtraction.

The spectra were processed using the standard procedure of the IRAF $Gemini$ package based on the PyRAF Python-based interface. Exposures from the same nodding position were added to boost the signal-to-noise ratio, then the sum of exposures from two different nodding positions were subtracted to remove background noise. Wavelength calibration was done against an Argon lamp in order to assign wavelength values to the observed pixels. 

Spectra of telluric standard stars with $T_{\rm eff} \sim 9700$~K were taken immediately before or after the science exposures to remove telluric absorption features in the quasars' observed spectra. These spectra were processed in a similar fashion, followed by a removal of the stars' intrinsic hydrogen absorption lines by fitting a Lorentzian profile to each hydrogen absorption line, and interpolating across this feature to connect the continuum on each side of the line. The quasars' spectra were divided by the corrected stellar spectra. The corrected quasar spectra were then multiplied by an artificial blackbody curve with a temperature corresponding to the telluric standard star, which yielded a cleaned, observed-frame quasar spectrum.

Flux calibrations were obtained by taking the Wide-field Infrared Survey Explorer \citep[WISE;][]{2010AJ....140.1868W} $W1$-band (at $3.4$~$\mu$m) apparent magnitudes, reported by SDSS Data Release 16 \citep{2020ApJS..250....8L}, and the $W1$~isophotal flux density $F_{\lambda}$(iso) given in Table 1 of \citet{2011ApJ...735..112J}. Flux densities at 3.4~$\mu$m were derived according to:
\begin{equation}
    \label{eq:fc}
    F_\lambda (3.4~\mu m) = F_{\lambda}(\rm iso) \cdot 10^{-mag/2.5}.
\end{equation} 
The flux densities at $3.4~\mu$m were extrapolated to flux densities at $1.63~\mu$m, roughly corresponding to \hbox{$\lambda_{\rm rest} = 5100~$\AA}, assuming an optical continuum of the form \hbox{$F_\nu \propto \nu^{-0.5}$} \citep[e.g.,][]{2001AJ....122..549V}.

\begin{figure*}
    \centering
    \includegraphics[width=0.8\linewidth]{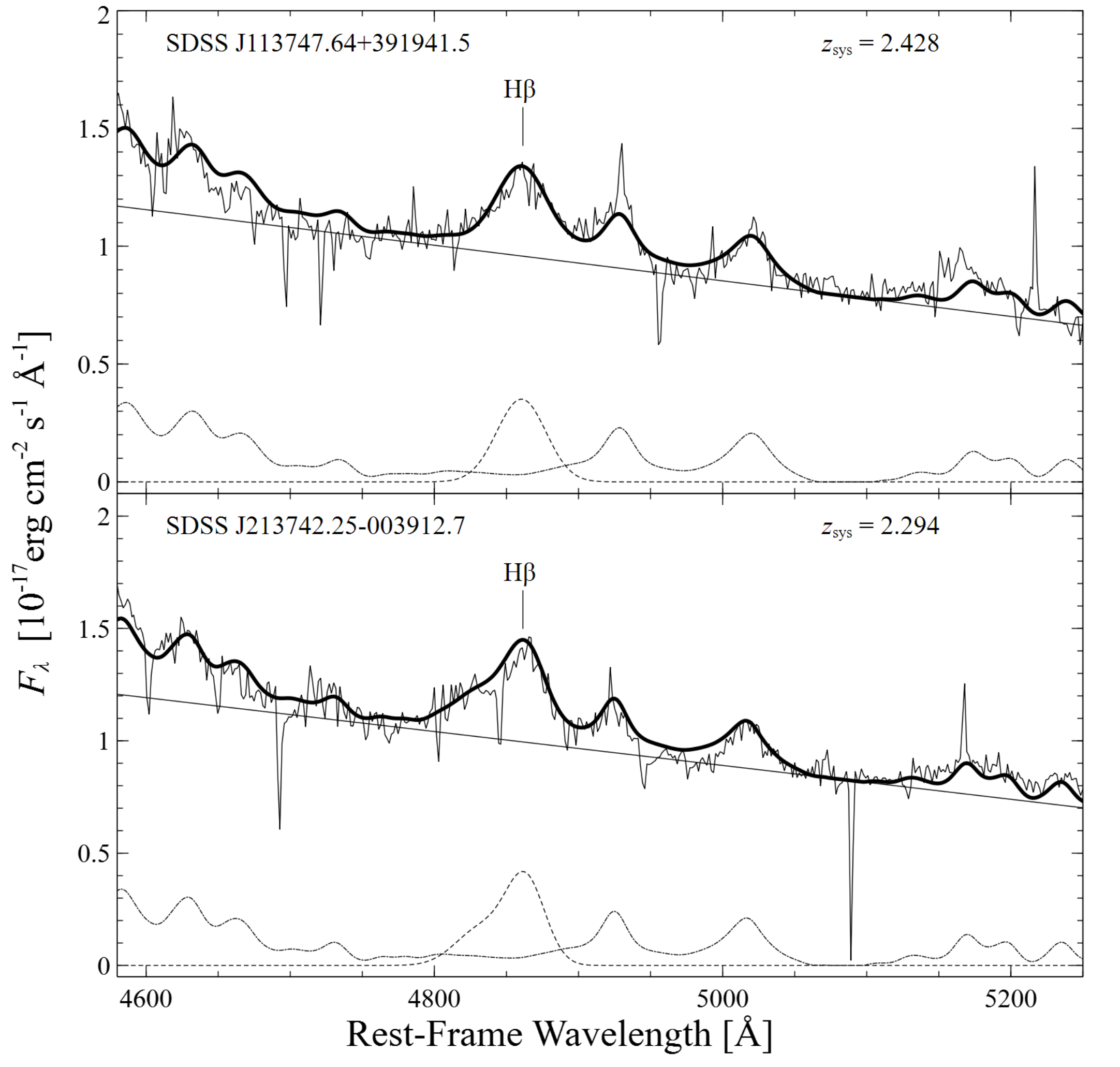}
    \caption{The NIR spectra of SDSS J1137$+$3919 (top) and SDSS J2137$-$0039 (bottom). In each panel, the continuous line is the observed spectrum of each quasar. The continuous straight line below the spectrum is the linear continuum fit. The dashed line is the H$\beta$ $\lambda$4861 profile modelled with two Gaussians. The dotted-dashed line is the \fetwo~ template from \citet{BG92}, which was broadened by 1500 \kms~for SDSS J1137$+$3919, and 1400 \kms~for J2137$-$0039. The bold solid line is the entire fitted spectrum.}
    \label{fig:spectra}
\end{figure*}

We modeled the spectra following the methods of \citet{Shemmer04}~and \citet{Shemmer10}. Our model consists of a linear continuum through the average flux densities of two narrow ($\sim$20~\AA) rest-frame bands centered on 4750~\AA~and 4975~\AA, a broadened \fetwo~ emission template \citep{BG92}, and two Gaussian profiles for the \hb~$\lambda$4861 emission-line. No [\othree] emission-lines are detectable in either spectrum, and we placed upper limits on their EWs by fitting a Gaussian feature where the [\othree] emission-lines should be such that they are indistinguishable from the continuum. The final, calibrated near infrared spectra of the two WLQs appear in Figure~\ref{fig:spectra}.

In both sources we detected weak and relatively narrow H$\beta$ lines as well as strong \fetwo~features compared to quasars at similar luminosities and redshifts \citep[e.g., see][]{Netzer07,2016ApJ...817...55S}. We also determined the systemic redshift ($z_{\rm sys}$) values from the observed-frame wavelength of the peak ($\lambda_{\rm peak}$) of the H$\beta$ emission-line, a similar treatment as in \citet{Matthews21} for sources that lack [\othree] emission. The $z_{\rm sys}$~values are larger than the redshifts reported by ~\citet{2020ApJS..250....8L} by $\Delta z$ = 0.008 in SDSS J1137$+$3919 and by \hbox{$\Delta z$ = 0.013} in SDSS J2137$-$0039, corresponding to velocity offsets (blueshifts) of 700 km~s$^{-1}$ and 1184 km~s$^{-1}$, respectively, which is consistent with typical velocity offsets between SDSS Pipeline redshifts and $z_{\rm sys}$~values observed in luminous, high-redshift quasars \citep[][Paper~I]{Dix20}. The rest-frame spectra in Figure~\ref{fig:spectra} have henceforth been corrected by $z_{\rm sys}$. Rest-frame EWs of \hb~$\lambda4861$, \fetwo~$\lambda\lambda4434-4684$, and the upper limit on the EWs of [\othree]~$\lambda5007$~were calculated for SDSS~J1137$+$3919 to be 16~\AA, 53~\AA, and $\leq 4$~\AA, and for SDSS J2137$-$0039 to be 20~\AA, 49~\AA, and $\leq 5$~\AA, respectively. The flux densities at a rest-frame wavelength of 5100~\AA~are $7.77 \times 10^{-18}$~\,ergs\,cm$^{-2}$\,s$^{-1}$\,\AA$^{-1}$ and $8.18 \times 10^{-18}$~\,ergs\,cm$^{-2}$\,s$^{-1}$\,\AA$^{-1}$, respectively.

\begin{deluxetable*}{lccccc}\label{table:obs}
\tablecolumns{6}
\tablewidth{0pc}
\tablecaption{\geminiN\ GNIRS $H$-Band Observation Log}
\tablehead{
\colhead{Quasar} &
\colhead{$z$\tablenotemark{a}} &
\colhead{$z_{\rm sys}$\tablenotemark{b}} &
\colhead{$\log \nu L_{\nu}$$(5100$\,\AA)} &
\colhead{Observation} &
\colhead{Exp.~Time} \\
\colhead{(SDSS J)} &
\colhead{} &
\colhead{} &
\colhead{(erg~s$^{-1}$)} &
\colhead{Dates} & 
\colhead{(s)}
}
\startdata
\hbox{113747.64$+$391941.5} & 2.420 & 2.428 & 45.8 & 2014 Mar 14, 20 &  7140 \\
\hbox{213742.25$-$003912.7} & 2.281 & 2.294 & 45.8 & 2014 Jun 29, Aug 04 &  7040
\enddata
\tablenotetext{a}{Obtained from visually-inspected redshifts (zvis) reported in SDSS Data Release~16~\citep{2020ApJS..250....8L}}
\tablenotetext{b}{Systemic redshifts (see \S~\ref{sec:spectroscopy} for details).}
\end{deluxetable*}

\end{document}